\documentclass{./iopjournal}

\usepackage[english]{babel}
\usepackage{graphics}
\usepackage{graphicx}
\usepackage{bm}
\usepackage{verbatim}
\usepackage{amssymb}
\usepackage{braket}
\usepackage{color}
\usepackage{wrapfig}
\usepackage{pifont}
\usepackage{amsfonts}
\usepackage{mathrsfs}
\usepackage{slashed}
\usepackage{amsmath}
\usepackage{epsfig}
\usepackage{latexsym}
\usepackage{psfrag}
\usepackage{nicefrac}
\usepackage{ragged2e}
\usepackage{etoolbox}
\usepackage{tikz}
\usepackage{pgfpages}
\usepackage{dsfont}
\usepackage{mathtools}
\usepackage{bbm}
\usepackage{cite}

\def\dblone{\hbox{$1\hskip -1.2pt\vrule depth 0pt height 1.6ex width 0.7pt
     \vrule depth 0pt height 0.3pt width 0.12em$}}
     \def\chiral4lo{$\mathrm{N}^4\mathrm{LO}$}

\begin{document}

\articletype{Paper}

\title{A Phenomenological Extension for Microscopic Optical Potentials}

\author{Matteo Vorabbi$^1$\orcid{0000-0002-1012-7238}, Michael Gennari$^{2}$\orcid{0000-0001-5271-8784}, Paolo Finelli$^{3,4}$\orcid{0000-0002-9958-993X}, Carlotta Giusti$^{5}$\orcid{0000-0003-1901-0885} and Petr Navr\'atil$^{6,7}$\orcid{0000-0002-6535-2141}}

\affil{$~^{1}$ Department of Physics, University of Surrey, Guildford, GU2 7XH, United Kingdom
}

\affil{$~^{2}$ Institut f\"ur Kernphysik and PRISMA++ Cluster of Excellence, Johannes Gutenberg-Universit\"at Mainz, 55128 Mainz, Germany
}

\affil{$~^{3}$ Dipartimento di Fisica e Astronomia, 
Universit\`{a} degli Studi di Bologna
}

\affil{$~^{4}$ INFN, Sezione di Bologna, Via Irnerio 46, I-40126 Bologna, Italy
}

\affil{$~^{5}$ INFN, Sezione di Pavia,  Via A. Bassi 6, I-27100 Pavia, Italy
}

\affil{$~^{6}$  TRIUMF, 4004 Wesbrook Mall, Vancouver, British Columbia, V6T 2A3, Canada
}

\affil{$~^{7}$  University of Victoria, 3800 Finnerty Road, Victoria, British Columbia V8P 5C2, Canada
}

\email{paolo.finelli@unibo.it, m.vorabbi@surrey.ac.uk}\\


\begin{abstract} 
Microscopic optical potentials constructed from realistic nucleon–nucleon interactions via multiple-scattering theory provide a first-principles description of nucleon–nucleus scattering. Nevertheless, such approaches often neglect medium corrections beyond Pauli blocking and fail to fully capture higher-order scattering contributions, leading to systematic under-prediction of absorption and deficiencies in angular distributions at low and intermediate energies. In this work we introduce a phenomenological correction scheme with an energy-dependent term designed to mimic correlation effects, dispersive contributions, and multi-step scattering processes. The correction is implemented in a minimal form to preserve the predictive character of the underlying microscopic model, while enabling improved flexibility in describing experimental observables. Applications to proton and neutron elastic scattering on light-mass nuclei demonstrate that the modified potentials yield enhanced agreement with measured differential cross sections, without sacrificing the microscopic foundation. This approach provides a practical pathway for incorporating missing medium and higher-order effects into optical model analyses relevant for nuclear structure and reaction studies.
\end{abstract}


\section{Introduction}

The optical model has long been a cornerstone in the theoretical description of nucleon–nucleus and nucleus–nucleus scattering. In its most general form, the optical potential represents the effective interaction between a projectile and a target nucleus, incorporating both elastic and inelastic channels through complex potential terms \cite{feshbach1992theoretical, goldberger2004collision}. Early optical model parametrizations were of purely phenomenological nature, designed to reproduce experimental cross sections and angular distributions across a broad energy range \cite{satchler1983direct, FOLDY1969447}. While such phenomenological potentials have proven to be highly successful in practice, their limited microscopic foundation reduces their predictive power, particularly when extrapolated to exotic nuclei and energies outside the range of fitted data.

To address this limitation, considerable effort has been devoted to the determination of microscopic optical potentials from nucleon–nucleon (NN) interactions constrained by scattering data and many-body theory. Approaches based on multiple-scattering expansions, Green’s function formalism, or effective interaction methods have yielded significant progress, providing a direct link between NN forces and reaction observables. These microscopic models incorporate some important effects, such as single-particle structure, Pauli blocking, or the density dependence of the nuclear medium, thereby grounding the optical potential in the nuclear structure theory. 

Recent efforts employ chiral effective field theory (EFT) to derive NN and three-nucleon (3N) forces consistent with underlying QCD symmetries to be employed in some computational framework.

Despite these advances, aligning fully microscopic potentials with experimental data remains a challenge. Areas requiring further development include the precise modelling of absorption at intermediate energies, as well as the reproduction of intricate fine structures in angular distributions and polarization observables \cite{Hebborn:2022vzm, Burrows:2020qvu, 10.3389/fphy.2022.1071971, PhysRevC.110.034605, Arellano:1995zz, PhysRevC.84.034606, qxj2-zbcn, Crespo:1992zz}.

Recent studies, like those performed by Vorabbi {\it et al.} 
\cite{Vorabbi:2015nra, Vorabbi:2017rvk, Gennari:2017yez, Vorabbi:2018bav, Vorabbi:2019ciy, Vorabbi:2020cgf, Vorabbi:2021kho, Vorabbi:2023mml, Vorabbi:2025idm}, combine ab initio No-Core Shell Model (NCSM) \cite{Navratil:2009ut} or self-consistent Green’s function (SCGF) \cite{PhysRevC.101.014318}  target densities with NN transition potentials folded within Watson's multiple-scattering framework. Their microscopic optical potentials, applied to light and medium-mass isotopes like carbon, oxygen, calcium and nickel, accurately reproduce elastic-proton differential cross sections and analyzing powers in the 65-250 MeV energy interval. At lower energies the quality of the agreement between theoretical predictions and empirical data degrades.

The origin of this deficiency can be traced back to missing medium and higher-order scattering corrections. Standard formulations truncate the multiple-scattering series at lowest non-trivial order and neglect dynamical correlations beyond the independent-particle approximation. As a result, important mechanisms, such as dispersive contributions from virtual excitations or multi-step scattering processes, are not fully incorporated \cite{Hebborn:2022vzm, Crespo:1991zz, PhysRevC.52.1992, Chinn:1993zz}.

A possible strategy to overcome these shortcomings is to introduce phenomenological corrections to otherwise microscopic potentials. Such corrections aim to mimic the impact of higher-order scattering and medium effects in a controlled and minimal fashion, preserving the predictive microscopic framework while enhancing quantitative agreement with experiment. The success of this approach depends critically on maintaining a balance, i.e., the corrections must be flexible enough to capture the missing physics and, at the same time, constrained enough to avoid degeneration with purely empirical parametrizations.

In this work, we propose a phenomenological correction scheme that improves microscopic optical potentials with an additional energy-dependent term. The form of the correction is guided by physical considerations, ensuring that it reflects plausible nuclear medium effects rather than arbitrary fitting. We apply this framework to nucleon-nucleus (NA) scattering on light-mass systems, comparing theoretical predictions with elastic cross sections. The results demonstrate that the corrected microscopic potentials yield improved agreement with data while retaining their theoretical foundation.

This paper is organized as follows. In Sec. 2, we briefly review the construction of microscopic optical potentials and the theoretical motivations for our correction scheme. Sec. 3 presents the formal definition of the phenomenological correction and discusses its physical interpretation. The determination of the correction factor chosen for the calculations is presented in Sec. 4. Numerical results for elastic neutron and proton scattering off $^{12}$C and $^{16}$O nuclei are shown in Sec. 5, where we benchmark the corrected potentials against experimental observables. Finally, Sec. 6 summarizes the conclusions and outlines directions for future work, including applications to exotic nuclei and reactions of astrophysical relevance.

\section{The microscopic optical potential}

In the next subsection we briefly summarize the main steps of the formal derivation of the optical potential following 
Ref.~\cite{Vorabbi:2015nra} and references therein.

\subsection{The spectator expansion}

The commonly used starting point for the description of the elastic scattering of a single projectile from a target of $A$ nucleons is the separation of the well--known Lippman-Schwinger (LS) equation for the transition operator $T$
\begin{equation}\label{generalscatteq}
T = V + V G_0 (E) T \, 
\end{equation}
into two parts,  i.e., an integral equation for $T$
\begin{equation}\label{firsttamp}
T = U + U G_0 (E) P T \, ,
\end{equation}
by the introduction of an optical potential operator $U$ that satisfies the following  integral equation
\begin{equation}\label{optpoteq}
U = V + V G_0 (E) Q U \, .
\end{equation}
In the above equations the operator $V$ represents the external interaction, such that the Hamiltonian for the entire $(A+1)$-particle
system is given by
\begin{equation}
H_{A+1} = H_0 + V \, .
\end{equation}
Asymptotically, the system is in an eigenstate of $H_0$, and the corresponding free propagator $G_0 (E)$ for the
projectile plus target nucleus system is
\begin{equation}
G_0 (E) = \frac{1}{E - H_0 + i \epsilon} \, .
\end{equation}
The projection operators $P$ and $Q$ have been introduced as follows
\begin{equation}
P + Q = \dblone \, ,
\end{equation}
where $P$ is taken to project onto the elastic
channel, such that, among other properties, we have
\begin{equation}\label{procommutator}
[G_0 , P] = 0 \, .
\end{equation}
For the scattering of a single particle projectile from an $A$-particle target the free Hamiltonian is given by
\begin{equation}
H_0 = h_0 + H_A \, ,
\end{equation}
where $h_0$ is the kinetic energy operator for the projectile and $H_A$ stands for the target Hamiltonian. Thus the projector $P$ can
be defined as
\begin{equation}
P = \frac{\ket{\Phi_A} \bra{\Phi_A}}{\braket{\Phi_A|\Phi_A}} \, ,
\end{equation}
where $\ket{\Phi_A}$ corresponds to the ground state of the target, satisfying the condition given in Eq.~(\ref{procommutator}), and
fulfilling
\begin{equation}
H_A \ket{\Phi_A} = E_A \ket{\Phi_A} \, .
\end{equation}
With these definitions, the transition operator for elastic scattering may be defined as $T_{\mathrm{el}} = PTP$, in which case
Eq.~(\ref{firsttamp}) can be written as
\begin{equation}\label{elastictransition}
T_{\mathrm{el}} = P U P + P U P G_0 (E) T_{\mathrm{el}} \, .
\end{equation}
Thus the transition operator for elastic scattering is given by a straightforward one-body integral equation, which requires, of course,
the knowledge of the operator $P U P$. The theoretical treatment we are going to introduce consists of a different formulation of the many-body equation,
Eq.~(\ref{optpoteq}), such as where $P U P$ can be calculated with reasonable accuracy without solving the
complete many-body problem. For simplicity we only assume the presence of two-body forces (but the formalism could be easily extended to three-body forces, as shown in 
Ref.~\cite{Vorabbi:2020cgf}). With this assumption, the operator $U$ for the optical potential can be expressed as
\begin{equation}
U = \sum_{i=1}^A U_i
\end{equation}
where $U_i$ is given by
\begin{equation}\label{optpoteq2}
U_i = v_{0i} + v_{0i} G_0 (E) Q \sum_{j=1}^A U_j \, ,
\end{equation}
provided that
\begin{equation}
V = \sum_{i=1}^A v_{0i} \, .
\end{equation}
The two-body potential $v_{0i}$ acts between the projectile and the {\it i}th target nucleon. Through the introduction of an effective scattering 
operator $\tau_i$ which satisfies
\begin{equation}\label{firstordertau}
\tau_i = v_{0i} + v_{0i} G_0 (E) Q \tau_i \, ,
\end{equation}
we can rearrange Eq.~(\ref{optpoteq2}) as
\begin{equation}
U_i = \tau_i + \tau_i G_0 (E) Q \sum_{j\neq i} U_j \, .
\end{equation}
This rearrangement process can be continued for all $A$ target particles, so that the operator for the optical potential can be
expanded in a series of $A$ terms of the form
\begin{equation}\label{spectatorexp}
U = \sum_{i=1}^A \tau_i + \sum_{i,j\neq i}^A \tau_{ij} + \sum_{i,j\neq i,k\neq i,j}^A \tau_{ijk} + \cdots \, .
\end{equation}
Eq.~(\ref{spectatorexp}) defines the so-called spectator expansion that derives its name
from the underlying idea that in lowest order all target constituents but the initially struck one (particle {\it i}) are ``passive''.
It is important to remark that the finite series given in Eq.~(\ref{spectatorexp}), together with the definitions of $\tau_i$, $\tau_{ij}$, and so on, is not unique,  due to the
many-body nature of $G_0 (E)$.

The aim of our manuscript is to propose a simple procedure to overcome the limitations of the lowest first-order approximation, adopted in Refs.~\cite{Vorabbi:2015nra, Vorabbi:2017rvk, Gennari:2017yez, Vorabbi:2018bav, Vorabbi:2019ciy, Vorabbi:2020cgf, Vorabbi:2021kho, Vorabbi:2023mml, Vorabbi:2025idm} to derive microscopic optical  potentials,
 avoiding complicated second-order calculations of the spectator expansion in Eq.~(\ref{spectatorexp}) (about whose convergence, in general, few assumptions have to be made).

\subsection{The first-order term}
The first-order term in the spectator expansion, $\tau_i$, as given by Eq.~(\ref{firstordertau}), is now examined, as outlined in Ref.~\cite{PhysRevC.52.1992}.
Since for elastic scattering $P \tau_i P$, or equivalently $\braket{\Phi_A |\tau_i |\Phi_A}$, is the only quantity that needs to be considered, Eq.~(\ref{firstordertau})
can be re-expressed as
\begin{equation}\label{hattaueq}
\tau_i = \hat{\tau}_i - \hat{\tau}_i G_0 (E) P \tau_i \; ,
\end{equation}
where $\hat{\tau_i}$ is defined as the solution of
\begin{equation}\label{hattaueq3}
\hat{\tau}_i = v_{0i} + v_{0i} G_0 (E) \hat{\tau}_i \, .
\end{equation}
Since Eq.~(\ref{hattaueq}) is a simple one-body integral equation, the main problem is to find a solution of Eq.~(\ref{hattaueq3}).
Due to the many-body character of $G_0 (E)$, Eq.~(\ref{hattaueq3}) is a many-body integral equation, and finding its solution is as difficult as
finding a solution for the original Eq.~(\ref{generalscatteq}). However, $G_0 (E)$ may be written as
\begin{align}
G_0 (E) &= \frac{1}{E - h_0 - H_A + i \epsilon} \\
&= \frac{1}{E - h_0 - h_i - W_i - H^i + i \epsilon} \, ,
\end{align}
where $h_0$ is the kinetic energy of the projectile, $h_i$ the kinetic energy of the {\it i}th target particle,
\begin{equation}
W_i = \sum_{j \neq i} v_{ij} \, ,
\end{equation}
and
\begin{equation}
H^i = H_A - h_i - \sum_{j \neq i} v_{ij} \, .
\end{equation}
Since $H^i$ has no explicit dependence on the {\it i}th particle, Eq.~(\ref{hattaueq3}) may be simplified by the replacement of
$H^i$ by an average energy $E^i$.
Thus, we consider $G_0 (E)$ to be $G_i (E)$, where
\begin{equation}\label{one_body_propagator}
G_i (E) = \frac{1}{(E-E^i) - h_0 - h_i - W_i + i \epsilon} \, ,
\end{equation}
so that $\hat{\tau}_i = \tilde{\tau}_i + (\mbox{higher-order corrections})$, and Eq.~(\ref{hattaueq3}) reduces to
\begin{equation}\label{tildetaueq}
\tilde{\tau}_i = v_{0i} + v_{0i} G_i (E) \tilde{\tau}_i \, .
\end{equation}
Equation~(\ref{tildetaueq}) can also be re-expressed as
\begin{equation}\label{eq_tau_tilde}
\tilde{\tau}_i = t_{0i} + t_{0i} g_i W_i G_i (E) \tilde{\tau}_i \, ,
\end{equation}
where the operators $t_{0i}$ and $g_i$ are defined as
\begin{equation}
t_{0i} = v_{0i} + v_{0i} g_i t_{0i}
\end{equation}
and
\begin{equation}
g_i = \frac{1}{(E-E^i) - h_0 - h_i + i \epsilon} \, .
\end{equation}
The quantity $W_i$ represents the coupling of the struck target nucleon to the residual nucleus. In practical calculations, $W_i$
has been taken to be an average one-body potential independent of the particle label {\it i}. It has been also used an average value of
the energy $E^i$, which has been taken equal to zero.
If we neglect the interaction of the struck target nucleon with the residual nucleus, i.e., we set $W_i = 0$, then we obtain $\hat{\tau} \approx \tilde{\tau} \approx t_{0i}$,
that corresponds to the impulse approximation (IA) to the optical potential adopted in Refs.~\cite{Vorabbi:2015nra, Vorabbi:2017rvk, Gennari:2017yez, Vorabbi:2018bav, Vorabbi:2019ciy, Vorabbi:2020cgf, Vorabbi:2021kho, Vorabbi:2023mml, Vorabbi:2025idm} to derive our microscopic optical potentials, where the operator $t_{0i}$ can be identified with the free NN $t$ matrix.

\subsection{Higher-order corrections}

When the coupling potential $W_i$ is included in Eq.~(\ref{eq_tau_tilde}), higher-order corrections, i.e., medium effects and rescattering terms, are included in the optical potential. A possible method to include such effects was derived in Refs.~\cite{PhysRevC.52.1992, Chinn:1993zz} and here we briefly review how formally this can be done.
We rewrite Eq.~(\ref{one_body_propagator}) as
\begin{equation}\label{relation_propagators}
G_i (E) = g_i + g_i W_i G_i (E) \, ,
\end{equation}
and with Eq.~(\ref{relation_propagators}) we can rewrite Eq.~(\ref{eq_tau_tilde}) as
\begin{equation}
\tilde{\tau}_i = t_{0i} + t_{0i} g_i \mathcal{T}_i g_i \tilde{\tau}_i = t_{0i} + w_{0i} g_i \tilde{\tau}_i \, ,
\end{equation}
where the operator $\mathcal{T}_i$ is the transition operator corresponding to the internal target potential $W_i$ and satisfies
\begin{equation}
\mathcal{T}_i = W_i + W_i g_i \mathcal{T}_i \, ,
\end{equation}
and the operator $w_{0i}$ is defined as follows
\begin{equation}\label{def_operator_w}
w_{0 i} \equiv t_{0i} g_i \mathcal{T}_i \, .
\end{equation}
To include medium effects we could guess to add corrections to the free $t$ matrix and express the matrix $\tilde{\tau}_i$ as
\begin{equation}
\tilde{\tau}_i = t_{0i} + \Delta_{0i} \, ,
\end{equation}
where
\begin{equation}
\Delta_{0 i} = \eta_{0 i} g_i t_{0 i} \, ,
\end{equation}
and
\begin{equation}
\eta_{0 i} = w_{0 i} + w_{0 i} g_i \eta_{0 i} \, .
\end{equation}
This method offers a clear way to include medium effects, but brings some challenges. First, the calculation of Eq.~(\ref{def_operator_w})
is not straightforward and it usually requires an additional approximation that was proved to be reasonable only for heavy targets \cite{Chinn:1993zz}. Second, this scheme requires the knowledge of the potential $W_i$ that, at least in principle, needs to be computed using the same approach that is used to calculate the target density. 
Nonetheless, we believe that the previous equations could be used as a starting point for a phenomenological modellization of, in a broader sense, higher-order effects like medium corrections or secondary scattering events. 

\section{Phenomeneological ansatz for higher-order effects}

Instead of following the aforementioned procedure, we propose a simpler phenomenological correction to include higher-order effects
related to the fact that the target nucleus is composed of nucleons that, in some way, all participate in
the scattering process. 
We are aware of the limitations of a phenomenological correction scheme, aim of our present manuscript is to propose a first simpler suggestion before more rigorous and complicated microscopic approaches are developed.  
In the following we provide a reasonable justification to our proposal.

We start considering the term $g_i W_i G_i (E)$ in Eq.~(\ref{eq_tau_tilde}), that represents an approximation of the full
propagator containing Pauli blocking and self-energy corrections.
Our goal is to adopt the approximation $\hat{\tau} \approx \tilde{\tau}$ and then find a way to calculate $\tilde{\tau}$ going beyond the IA. 

A natural path would be solving Eq.~(\ref{eq_tau_tilde}), that requires the evaluation of the effective {\it in-medium} propagator
\begin{equation}\label{in_medium_propagator}
g_m \equiv g_i W_i G_i (E) \, ,
\end{equation}
and, ultimately, the knowledge of the potential $W_i$. Instead of pursuing this path we propose an alternative computational scheme.

The starting point is to explore the effect of the operator $W_i G_i (E)$ in Eq.~(\ref{eq_tau_tilde}), which describes the medium modification due to the coupling
of the struck nucleon in the target to the residual nucleus. This was already investigated in Ref.~\cite{PhysRevC.52.1992, Chinn:1993zz} using a mean field potential. 

We propose a different strategy. Our basic assumption is to approximate the medium effects entering Eq.~(\ref{eq_tau_tilde}) with a scaling factor $\lambda (E)$, dependent only on the projectile energy, that mimics the medium effects simply suppressing the strength of the free propagator $g_i$. 
This idea relies on the existing literature about the influence of higher-order effects on experimental observables. 
In-medium cross sections are usually reduced compared to the free-space case \cite{JEUKENNE197683,Li:1993rwa}. In fact, when nucleons (or other particles) are inside a dense medium (like nuclear matter), several effects modify their scattering properties compared to scattering in free space, among them Pauli blocking and in-medium effective masses.
As a consequence, we propose an ansatz based on the assumption that the transition amplitudes must be reduced. This leads us to propose a minimal correction like 
\begin{equation}\label{medium_effects_approximation}
g_i W_i G_i (E) \approx - \lambda (E) \, g_i \, .
\end{equation}
With this choice, Eq.~(\ref{eq_tau_tilde}) is approximated as
\begin{equation}\label{eq_tau_tilde_ansatz_1}
\tilde{\tau}_i \approx \sum_{n=0}^{N_{{\rm tr}}} (-1)^n \lambda^n (E) \, ( t_{0i} g_i )^n t_{0i} \, ,
\end{equation}
where the series necessarily must be truncated at some order $N_{{\rm tr}}$ (since convergence is not clearly assured).
With this assumption the in-medium two-nucleon scattering operator $\tilde{\tau}$ is described as a series of
two-nucleon multiple-scattering processes that are systematically suppressed by the factor $\lambda (E)$.
It is worth noting that the leading term of Eq.~(\ref{eq_tau_tilde_ansatz_1}) is the free NN $t$ matrix used in our previous work based on the impulse approximation, that is known to be a valid approximation to the operator $\tilde{\tau}$ at high energies.

Within this computational scheme, the final expression of the optical potential is the same as our previous work (see, for instance, Ref.~\cite{Vorabbi:2020cgf}), with the only difference that the free $t$ matrix is replaced by the in-medium $\tilde{\tau}$ matrix approximated by the expansion in Eq.~(\ref{eq_tau_tilde_ansatz_1}), i.e.,
\begin{equation}\label{optical_potential_formula}
\begin{split}
U_{\mathrm{el}}^{\textbf p} ({\bm q} , {\bm K}) &= \sum_{\mathcal{N} =p,n} \int d {\bm P} \; \eta ({\bm q} , {\bm K} , {\bm P}) \;
\tilde{\tau}_{{\textbf p} \mathcal{N}} \left[ {\bm q} , \frac{1}{2} \left( \frac{A+1}{A} {\bm K} + \sqrt{\frac{A-1}{A}} {\bm P} \right) ; E \right] \\
&\times \rho_{\mathcal{N}} \left( {\bm P} + \frac{1}{2} \sqrt{\frac{A-1}{A}} {\bm q} , {\bm P} - \frac{1}{2} \sqrt{\frac{A-1}{A}} {\bm q} \right) \, ,
\end{split}
\end{equation}
where ${\textbf p}$ identifies the projectile (either a proton $p$ or a neutron $n$), ${\bm q} = {\bm k}^{\prime} - {\bm k}$ is the momentum transfer,
${\bm K} = ( {\bm k}^{\prime} + {\bm k} ) / 2$ is the average momentum, and ${\bm P}$ is an integration variable. The variables ${\bm k}$ and ${\bm k}^{\prime}$ used to
define the vectors entering Eq.~(\ref{optical_potential_formula}) are the initial and final momenta of the projectile nucleon in the NA center of mass frame.
The quantity $\rho_{\mathcal{N}}$ is the one-body nonlocal density of the target, computed with the NCSM, and $\eta$ is the M\o ller factor that imposes the Lorentz invariance when the $\tilde{\tau}$ matrix is transformed from the NN frame, where it is evaluated, to the NA frame, where it is needed for the calculation of the potential.
The optical potential of Eq.~(\ref{optical_potential_formula}) represents the input quantity that is needed to solve Eq.~(\ref{elastictransition}), from which we can obtain
the scattering amplitude and consequently the scattering observables, such as the differential cross section and the analyzing power.

Before proceeding with our ansatz some considerations are in order.

%

We are aware that replacing the operator combination $W_i G_i(E)$  in Eq.~(\ref{medium_effects_approximation}) by the scalar suppression factor $-\lambda(E)$ represents a strong simplification in our optical potential model. Even if this simplification can be physically motivated by the well known reduction of in-medium cross sections relative to the free-space case, it compresses a highly nontrivial set of medium effects into a single energy-dependent parameter. In general, medium corrections are expected to depend not only on the projectile energy, but also on quantities such as the local density or the Fermi momentum and the relative momentum of the interacting nucleons. In this work we limit ourselves to propose our ansatz, and we formulate it at a first simple level, but, of course, it could be improved assuming a more general formulation. 

Another critical aspect of the series in Eq.~(\ref{eq_tau_tilde_ansatz_1}) is that the proposed expansion is not controlled by a perturbative (or some small controllable) parameter. Therefore, the truncation at a finite order $N_{\rm tr}$ is not guaranteed to be systematically improvable, and the convergence properties of the expansion remain uncertain in the general case. Convergence of the expansion has to be tested case by case. The results presented in Sect.~\ref{results} show that only a relative small number of terms must be included.

One additional concern is that we are modifying the amplitude via a real scalar function $\lambda(E)$, i.e.,
we are changing both real and imaginary parts in the same way. This could lead to a small theoretical inconsistency since 
$\mathrm{Re}~ \tilde{\tau}(E)$ and $\mathrm{Im} ~\tilde{\tau}(E)$ should be related by  Kramers–Kronig–type relations in the energy domain. In the context of the  multiple scattering theory, optical potentials inherit part of the analytical structure of the underlying two-body scattering amplitude, but the approximations used in practical computations usually do not guarantee that the resulting optical potential completely satisfies the full dispersion relations. Since a real scaling factor generally violates dispersion relations, for future applications it would be interesting to explore a complex $\lambda(E)$ as initial ansatz.

Finally, our  choice to employ a minus sign in the expansion, see Eq.~(\ref{medium_effects_approximation}), leads to an expansion series of contributions with alternating signs. As a consequence, as $N_{\rm tr}$ increases, the {\it in-medium} contributions from high-order terms are somewhat mitigated, as reasonably expected.

\section{Determination of the expansion parameter $\lambda (E)$}
The immediate problem with this approach is that the parameter $\lambda (E)$ is not known a priori and it has to be determined.
Among many available choices we propose a solution in which $\lambda$ is somewhat related to the probability of an additional scattering event for the projectile nucleon off the target nucleus. We believe that 
the probability that the nucleon interacts with another nucleon in the target, between $s$ and $s + \Delta s$ along its trajectory inside the target volume, can be reasonably described by the following relations:
\begin{align}
P_{Np} = \sigma_{Np} (E) \, \rho_p (s) \, \Delta s \, , \label{diff_prob_pro} \\
P_{Nn} = \sigma_{Nn} (E) \, \rho_n (s) \, \Delta s \, , \label{diff_prob_neu}
\end{align}
where $N$ stands for proton ($p$) and neutron ($n$), $\sigma_{NN} (E)$ is the NN elastic cross section, and
$\rho_N (s)$ is the nuclear density at the position $s$ on the nucleon's trajectory.

In the calculations the step parameter $\Delta s$ has to be reasonably evaluated in comparison with the diameter of the target nucleus.
The density $\rho_N (r)$ is the diagonal part of the nonlocal density that is used in the folding integral to calculate the optical potential.
The NN elastic cross sections are computed with the same chiral interaction used to compute the $t$ matrix and the target density in the calculation of the optical potential.
For the proton-proton case, the calculation of the elastic cross section is done neglecting the Coulomb interaction, since we are considering that these effects take place at very short-range distances, where nuclear forces are supposed to dominate over the Coulomb interaction.

\begin{figure}
\centering
\includegraphics[scale=0.2, angle=0.0]{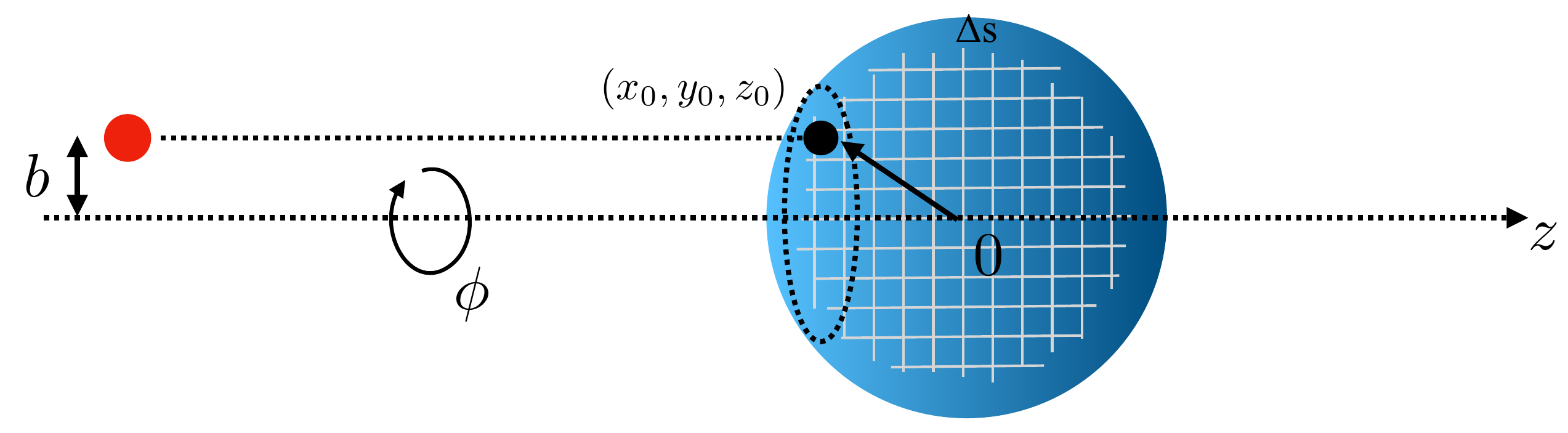}
\caption{\label{fig1}  Definition of the coordinate system and geometrical quantities for the secondary nucleon-nucleus scattering event. The point (($x_0$, $y_0$, $z_0$)) specifies the entrance position, with the $z$ axis aligned along the longitudinal direction, $b$ denotes the transverse impact parameter in the plane perpendicular to $z$, and $\Delta s$  indicates the infinitesimal trajectory step.
}
\end{figure}

The idea is to consider a nucleus centered at the origin and assume that the incoming nucleons are initially travelling coming from $- \infty$ parallel to the $\hat{z}$ axis,
as shown in  Fig.~\ref{fig1}. The shape of the nucleus is obtained from the nuclear density obtained from a NCSM approach.
We randomly generate an entry point $(x_0 , y_0 , z_0 )$ where the incoming nucleon initially collides against the nucleus. The prescription we adopt consists in generating a random value for the impact parameter $b$ as
\begin{equation}\label{impact_parameter}
b = \sqrt{n_r} \, R_{\mathrm{eff}} \, ,
\end{equation}
where $n_r$ is a random number generated between 0 and 1 and $R_{\mathrm{eff}}$ is the effective nuclear radius computed as
\begin{equation}
R_{\mathrm{eff}} = \sqrt{ \frac{Z \braket{r_p^2} + N \braket{r_n^2}}{Z + N}} \, ,
\end{equation}
where $\braket{r_p^2}$ and $\braket{r_n^2}$ are the proton and neutron mean squared radii.

The probability of selecting a particular impact parameter $b$ in a uniform two-dimensional distribution follows from $\Delta P \propto b \, \Delta b$.
To generate $b$ with this probability distribution we set
\begin{equation}
n_r = \frac{b^2}{R_{\mathrm{eff}}^2} \, , 
\end{equation}
that leads to Eq.~(\ref{impact_parameter}).

Once $b$ is generated, we generate another random number $n_r^\prime$ between 0 and 1 and we calculate the azimuthal angle $\phi$ as
\begin{equation}\label{first_sample_phi}
\phi = 2 \pi n_r^\prime \, .
\end{equation}
The components of the entry point are obtained as
\begin{align}
x_0 &= b \cos \phi \, , \\
y_0 &= b \sin \phi \, , \\
z_0 &= - \sqrt{R_{\mathrm{eff}}^2 - x_0^2 - y_0^2} \, .
\end{align}
For simplicity, we assume that the nucleon follows a linear trajectory inside the nucleus that we divide in steps of equal length $\Delta s$.
For each step $\Delta s$ along the trajectory we calculate the probabilities $P_{Np}$ and $P_{Nn}$ using Eqs.(\ref{diff_prob_pro}) and (\ref{diff_prob_neu}).  With this procedure we assume that the trajectory of the nucleon inside the nucleus continues to be parallel to the $z$ axis, meaning that the angle $\phi$ is the same
as that one obtained from Eq.~(\ref{first_sample_phi}), while $\theta = 0$. This assumption can be easily refined to keep into account a small component of transverse momentum.

After each probability $P_{Np}$ and $P_{Nn}$ has been computed, we generate a random number $\tilde{n}_r$ between 0 and 1 and we compare it with the sum $P_{Np} + P_{Nn}$. 
This represents the total probability of an interaction event.
If $\tilde{n}_r \ge P_{Np} + P_{Nn}$ no interaction occurred and we continue this process for the next step along the trajectory, until 
the scattering event happens.
In the case of a successful event, i.e., $\tilde{n}_r < P_{Np} + P_{Nn}$, we compute the normalised probability ratios $\bar{P}$, reducing the interaction problem to the problem of sampling an unknown distribution of secondary scattering events.
Then, generating another random number $\bar{n}_r$, we decide if the interaction took place with a neutron or a proton and store the information
in the {\it cumulative events} variable $I_i^{(NN)}$.

At this point, if we are only interested in the probability of at least one scattering, we can stop the algorithm and re-start it again for a different impact parameter
and a different entry point. We repeat the procedure for a number of trials $N_{\mathrm{c}}$
and the final probabilities, i.e., the $\lambda$ values that we are looking for, are given by
\begin{align}
\lambda_{pp} &= \frac{{\displaystyle \sum_{i=1}^{N_{\mathrm{c}}} b_i I_i^{(pp)}}}{{\displaystyle \sum_{i=1}^{N_{\mathrm{c}}} b_i}} \, , \\
\lambda_{pn} &= \frac{{\displaystyle \sum_{i=1}^{N_{\mathrm{c}}} b_i I_i^{(pn)}}}{{\displaystyle \sum_{i=1}^{N_{\mathrm{c}}} b_i}}  \, .
\end{align}

The procedure can be extended to calculate the probability of two or more scattering events, but we restrict the present analysis to a single additional scattering event. This choice is consistent with the theoretical framework in which the model is developed and the projectile energies considered.

This scheme may have some weaknesses, but our purpose  here is to find a compromise between physical intuition and simplicity.

A possible issue is that the scheme mixes a quantum multiple scattering framework with a semiclassical straight-line picture. The projectile is assumed to follow a linear trajectory through the nucleus, with no deflection until the additional interaction occurs, and with the probability of interaction sampled locally through the product of density and free-space elastic cross section. This is reminiscent of the  Glauber theory, but the optical potential under development is constructed within a microscopic quantum framework. In this sense, the present prescription produces a hybrid model.

There is also a possible ambiguity about the treatment of the additional scattering event. In fact, in the microscopic optical potential framework, the projectile is already interacting with the target through the folded first-order term, while the Monte Carlo algorithm estimates the probability of one more collision as it passes through the nucleus. The proposed ansatz, for which in the leading term of Eq.~(\ref{eq_tau_tilde_ansatz_1}) only the free NN $t$ matrix survives, seems to be consistent nonetheless with the correct bookkeeping of the spectator expansion.

From a broader perspective, the central weakness of the scheme is that it gives $\lambda$ an appealing probabilistic meaning, but that meaning is only loosely connected to the role $\lambda$ plays in the operator expansion for $\tilde\tau_i$. The Monte Carlo procedure estimates something like a classical path-integrated collision probability in a finite nucleus. The parameter in the optical potential ansatz, by contrast, controls the suppression of quantum multiple-scattering operator structures. The two ideas are not obviously equivalent. Thus, while the scheme is more physically motivated 
than taking $\lambda(E)$ as a free fit parameter and we believe that it may provide a plausible phenomenological estimate for the energy dependence of $\lambda$,  the connection between the quantities computed within this scheme and the role played by $\lambda$ in the operator expansion of 
$\tilde{\tau}_i$ remains largely heuristic. Our approach falls short of a genuine microscopic derivation of in-medium effects, it can rather be considered as an effective estimate of the likelihood of rescattering processes that are not explicitly included in the leading-order spectator expansion.

In several aspects our simple prescription could be improved, for instance, by employing an {\it in-medium} interaction instead of free-space NN elastic cross sections, or by improving the definition of the entry point, or the determination of the scatterer's position.

As a conclusion of this section, it is worth mentioning that the proposed approach, despite all its limitations, does not include any fitting  procedure whatsoever, it is self-consistent, and based only on the free NN interaction and the target density.

\subsection{Proposed algorithm}

In this subsection we collect all the necessary steps to implement our approach into a general scattering formula. In chronological order:

\begin{enumerate}

\item Generate randomly the impact parameter $b$ and the azimuthal angle $\phi$.
 
\item For each step $ \Delta s$ calculate the secondary scattering probability $P_{Np}$ and $P_{Nn}$.

\item Generate another random number to establish if a secondary scattering happened or not.

\item Calculate the normalised probability ratios if the interaction has happened.

\item Calculate the cumulative variables $I_i^{(pp)}$and  $I_i^{(pn)}$.

\item Repeat the above procedure with different impact parameter $b$ and entry point.

\item Collect all the results in the variables $ \lambda_{pp}$ and $\lambda_{pn}$ and generate the NN scattering matrix $\tilde{\tau}_i$ of Eq.~(\ref{eq_tau_tilde_ansatz_1}).

\end{enumerate}

\section{Theoretical predictions}

In this section we investigate the effects of the phenomenological correction proposed in the previous sections on the microscopic optical potentials derived in 
Refs.~\cite{Vorabbi:2015nra, Vorabbi:2017rvk, Gennari:2017yez, Vorabbi:2018bav, Vorabbi:2019ciy, Vorabbi:2020cgf, Vorabbi:2021kho, Vorabbi:2023mml, Vorabbi:2025idm}. Numerical results for the differential cross section and analyzing power of elastic neutron and proton scattering off $^{12}$C  and $^{16}$O are presented and discussed in comparison with available experimental data. 
  
Before presenting our numerical results it may be worth briefly discussing the main theoretical input of the calculations: the NN chiral potential,  which is the only theoretical input of our microscopic optical potentials and characterizes both the projectile-target interaction and the NCSM description of the nuclear targets.

\subsection{Theoretical input: the nuclear chiral potential}

In this manuscript we make exclusive use of the most recent generation of chiral interactions derived within the formalism of the Chiral Perturbation Theory (ChPT). 
Within this framework, the NN interaction is governed by the (approximate) chiral symmetry
of the low-energy realization of QCD. ChPT provides a description of nuclear systems in terms of single and multiple pion exchanges (long- and medium-range components) and contact interactions between the nucleons in order to
parametrize the short-range behavior. For all details, for more recent developments and interpretation we refer the reader to Refs.~\cite{ Epelbaum:2008ga, Machleidt:2011zz}. The free parameters of the theory are determined by reproducing data in the NN and 3N sectors.

In our previous work \cite{Vorabbi:2015nra, Vorabbi:2017rvk, Gennari:2017yez, Vorabbi:2018bav, Vorabbi:2019ciy, Vorabbi:2020cgf, Vorabbi:2021kho, Vorabbi:2023mml, Vorabbi:2025idm} we applied chiral NN potentials at N$^{3}$LO (next-to-next-to-next-to-leading order) and N$^{4}$LO (next-to-next-to-next-to-next-to-leading order), and at N$^{2}$LO (next-to-next-to-leading order) for the 3N sector.
At the moment, because of the high computational resources needed, it is impossible to achieve a full consistency between the NN potentials, employed for the target description and the elastic reaction process, and the inclusion of 3N forces.

Every calculation has been performed with the NN chiral interaction developed by Entem {\it et al.} \cite{PhysRevC.96.024004} up to the fifth order (N$^{4}$LO), with a
500 MeV cutoff and the three-nucleon local-nonlocal (3Nlnl) chiral interaction at N$^{2}$LO presented in Refs.~\cite{Navratil:2007zn,PhysRevC.101.014318},
with the low-energy constants fixed at $c_D = -1.8$ and $c_E =-0.31$ \cite{Gysbers2019}.
The same NN chiral interaction used to compute the nuclear density is consistently used in the calculation of the NN $t$ matrix.
For the two nuclei $^{12}$C and $^{16}$O considered, we employed $\hbar \omega = 18$ MeV and a $\lambda_{\mathrm{SRG}} = 1.8$ fm$^{-1}$ cutoff for the similarity renormalization
group (SRG) \cite{Bogner:2003wn, Bogner:2009bt} procedure (including the SRG induced 3N force in all the calculations), and
we performed calculations up to a maximum number of harmonic oscillator shells $N_{max} = 8$.
In the following, the interaction employed will be denoted as NN–N$^4$LO+3Nlnl.

\subsection{Results}
\label{results}

In our previous papers the optical potential was calculated at the first order of the spectator expansion as the folding integral  of the nuclear density of the target and the free NN $t$ matrix. With the proposed phenomenological correction, the optical potential is calculated from the expression in Eq.~(\ref{optical_potential_formula}), where the free NN $t$ matrix of our previous work is replaced by the sum in Eq.~(\ref{eq_tau_tilde_ansatz_1}), which in the calculations must necessarily be truncated at some order 
$N_{{\rm tr}}$. The first aim of the present calculations is to validate the phenomenological correction with respect to the main limitations of the sum, i.e., to establish at which order the sum should be truncated and test if some kind of convergence is reached.

The numerical results exhibit a clear and systematic improvement in comparison with experimental data with increasing truncation order $N_{{\rm tr}}$
and, at the same time, an acceptable degree of convergence. 
As a general statement, we can affirm that the comparison among successive orders indicates a well-behaved convergence pattern, with higher-order calculations 
($N_{{\rm tr}}=3,4$), yielding a stable and smooth description of the differential cross section across the full angular range, regardless of the target nucleus.

\begin{figure}
\centering
\includegraphics[scale=0.65, angle=0.0]{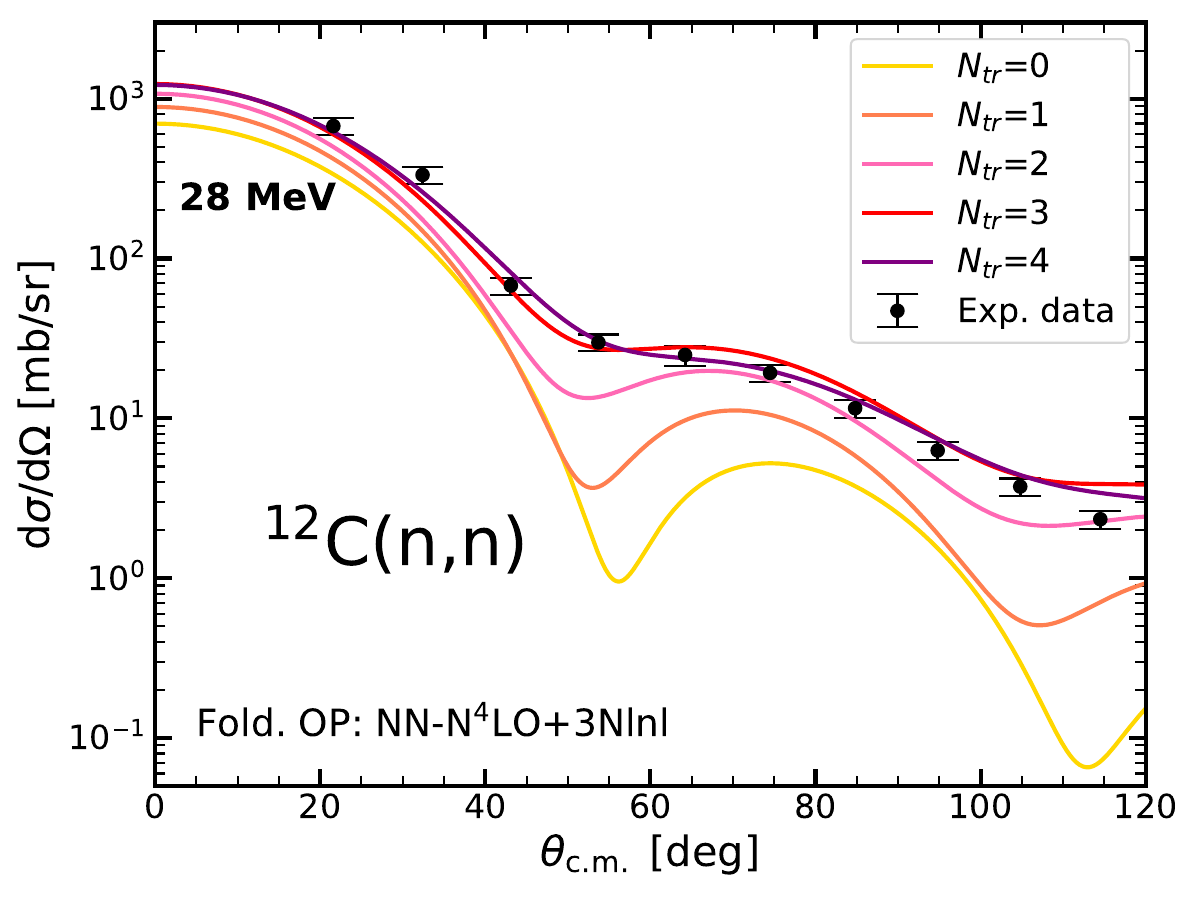}
\caption{\label{fig12c_28}  Differential cross sections as functions of the center-of-mass scattering angle for the reaction $^{12}$C($n,n$) at 28 MeV. The folded optical potential is calculated with 
the NN-N$^4$LO+3Nlnl chiral interaction. The results are shown for successive expansion orders: 
$N_{{\rm tr}}=0$ (yellow line), 1 (orange), 2 (pink), 3 (red), 4 (purple).
Experimental data (with error bars) are taken from Ref.~\cite{CHIBA1997305}.}
\end{figure}

This is clearly illustrated in Fig.~\ref{fig12c_28}, in which the differential cross section for the elastic reaction $^{12}$C($n,n$) at a projectile energy of 28 MeV is plotted as functions of the center-of-mass scattering angle $\theta_{\rm c.m.}$. Results are shown for successive expansion orders $N_{{\rm tr}}=0-4$. 

The systematic improvement with increasing order illustrates good convergence, with higher orders ($N_{{\rm tr}}=3$ and 4) providing a quantitatively reliable reproduction of the experimental diffraction pattern, both in terms of the position of the minima and the overall angular dependence. The improvement is especially notable when compared to the leading-order result, where the phenomenological correction is not applied, which significantly underestimates the cross section at large angles and fails to reproduce the structure of the diffraction minima. Of course, it could have expected that an optical potential based on the IA is unable to describe data at 28 MeV. The good agreement obtained at higher orders indicates that the combined use of the folded microscopic optical potential based on the NN–N$^{4}$LO+3Nlnl interaction and the present truncation scheme provides a physically consistent and numerically robust framework for describing elastic scattering observables even at this energy.

It is worth emphasizing that 28 MeV is an energy value close to the lower boundary of the typical applicability range of microscopic optical potential approaches and therefore represents a particularly stringent test of the present method. By exploring this regime, we aim to identify a truncation order $N_{\rm tr}$ that ensures reliable convergence even in conditions where the underlying assumptions on the nuclear structure and reaction dynamics are expected to become less accurate. In this sense, the present analysis provides a conservative benchmark for the selection of $N_{\rm tr}$ in more favourable kinematical regimes.

To further investigate if this agreement with empirical data holds consistently for different reactions at different energies, we extend our analysis to different energies and to elastic neutron and proton scattering on $^{12}$C and $^{16}$O.
To improve the clarity of our plots we generally display results only for $N_{{\rm tr}} =0$ and $3$.

\begin{figure}
\centering
\includegraphics[scale=0.65, angle=0.0]{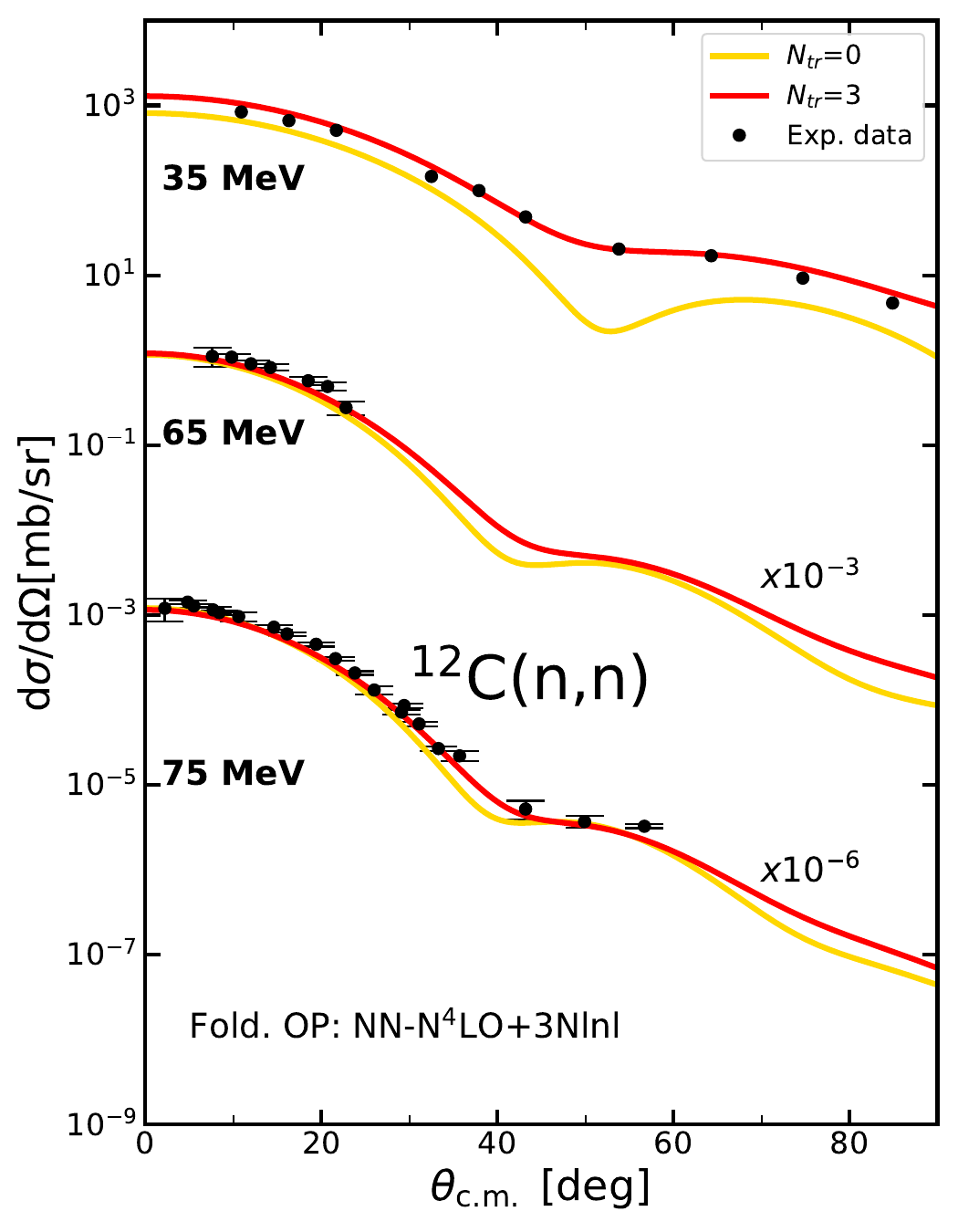}
\caption{\label{fign_12c} 
Differential cross sections as functions of the center-of-mass scattering angle for the reaction $^{12}$C($n,n$) at 35, 65 MeV (scaled by a factor $10^{-3}$) and 75 MeV (scaled by a factor $10^{-6}$). The folded optical potential is calculated with the NN-N$^4$LO+3Nlnl chiral interaction. The results are shown only for the orders $N_{{\rm tr}}=0$ (yellow lines) and $N_{{\rm tr}}=3$3 (red lines).
Experimental data (with error bars where available) are taken from Ref.~\cite{NIIZEKI1990455,Baba01082002,PhysRevC.70.054613}. }
\end{figure}

Fig.~\ref{fign_12c} shows the differential cross section for elastic neutron scattering on $^{12}\mathrm{C}$ for incident neutron energies of 35, 65, and 75 MeV.
Theoretical results are shown only for $N_{{\rm tr}}=0$ and $3$ and compared with available experimental data.
The comparison illustrates the energy dependence of the angular distributions and the improvement of the theoretical description with increasing model complexity, particularly in reproducing the magnitude and oscillatory structure of the measured cross sections.
In all three cases the leading-order calculation tends to underestimate the cross section at intermediate and large scattering angles and fails to reproduce the detailed structure of the angular distribution, particularly in the region of the diffraction minima. The discrepancies become increasingly pronounced as the scattering angle increases, highlighting the limitations of a first-order treatment.
The inclusion of higher-order contributions ($N_{\rm tr}=3$) significantly improves the agreement with experimental data across the full angular range. This is especially evident in the correct reproduction of the slope of the cross section and in the improved description of the diffraction pattern, both in terms of the position and depth of the minima. Although it is more evident at lower energies, the effect is consistently observed at all considered energies, indicating that the role of multiple-scattering corrections remains non-negligible even in regimes where the leading-order approximation is often assumed to be adequate.

It is also worth noting that the overall quality of the agreement improves with increasing projectile energy, as expected within the framework of our microscopic optical potentials. At 75 MeV the higher-order calculation provides an excellent description of the data over several orders of magnitude in the cross section, while at lower energies (e.g., 35 MeV) residual discrepancies at large angles persist, reflecting the increasing importance of mechanisms not fully accounted for within the present framework.

Overall, these results support the robustness of the proposed truncation scheme and confirm that the inclusion of a limited number of higher-order terms ($N_{\rm tr}\sim 3$) is sufficient to achieve a quantitatively reliable description of elastic scattering observables over a broad range of energies. We stress the non trivial aspect of the proposed approach that along the whole procedure no part of the model has been calibrated over any experimental data.

\begin{figure}
\centering
\includegraphics[scale=0.65, angle=0.0]{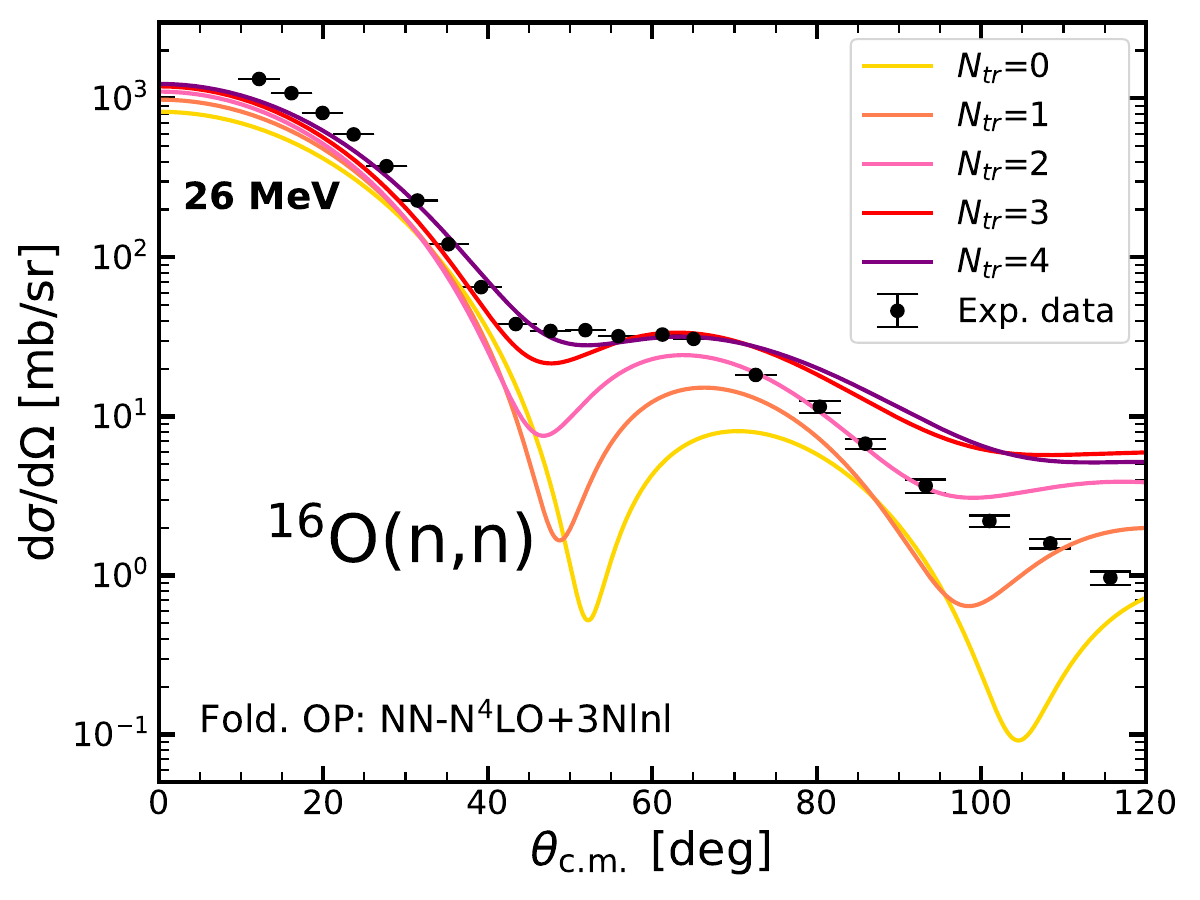}
\caption{\label{fig16o_26}  Differential cross sections as functions of the center-of-mass scattering angle for the reaction $^{16}$O($n,n$) at 26 MeV. The folded optical potential is calculated with 
the NN-N$^4$LO+3Nlnl chiral interaction. The results are shown for successive expansion orders: 
$N_{{\rm tr}}=0$ (yellow line), 1 (orange), 2 (pink), 3 (red), 4 (purple).
Experimental data (with error bars where available) are taken from Ref.~\cite{PhysRevC.33.1826}.}
\end{figure}

In Fig.~\ref{fig16o_26} the calculated differential cross section of the reaction $^{16}$O($n,n$) at 26 MeV is shown for successive expansion orders $N_{{\rm tr}}=0-4$ in comparison with available data. The results can be directly compared with the corresponding results of elastic neutron scattering on $^{12}$C at 28 MeV in Fig.~\ref{fig12c_28}.
Also in this case the inclusion of higher-order terms, with calculations at $N_{\rm tr}=3$ and 4, leads to a systematic improvement of the differential cross section, providing a stable result and a reasonable description of the angular distribution.
Also in this case the leading-order result ($N_{\rm tr}=0$) significantly underestimates the cross section at intermediate and large scattering angles and fails to reproduce the pronounced diffraction minimum around $\theta_{\mathrm{c.m.}}\sim 50^o$. The progressive inclusion of higher-order contributions results in a clear refinement of both the magnitude and the structure of the cross section, with $N_{\rm tr}=3$ already capturing the position and depth of the main minimum with good accuracy. The $N_{\rm tr}=4$ result confirms the stability of the expansion, with only marginal corrections relative to the previous order. The agreement with the data is slightly worse than that obtained at 28 MeV for $^{12}$C displayed in Fig.~\ref{fig12c_28}, suggesting that the lower energy limit of our ansatz has been reached.

The comparison between the results for $^{12}$C and $^{16}$O shows that,  despite the different nuclear structure and slightly different projectile energy, the convergence behaviour appears equally robust. This suggests that the truncation scheme is not strongly sensitive to the specific target nucleus and retains its effectiveness across systems with different density distributions. At the same time, the more pronounced diffraction pattern observed in $^{16}$O provides a more stringent test of the method, particularly in the region of the first minimum.

Overall, the results indicate that a truncation order of $N_{\rm tr}\sim 3$ is sufficient to achieve a quantitatively reliable description of elastic scattering observables in this energy regime. This reinforces the conclusion that the proposed expansion provides a consistent and transferable framework for incorporating higher-order effects in microscopic optical potential calculations and can be useful to extend to lower values the range of energies for which the multiple scattering formalism limited to the single-scattering approximation can be effectively used.

\begin{figure}
\centering
\includegraphics[scale=0.65, angle=0.0]{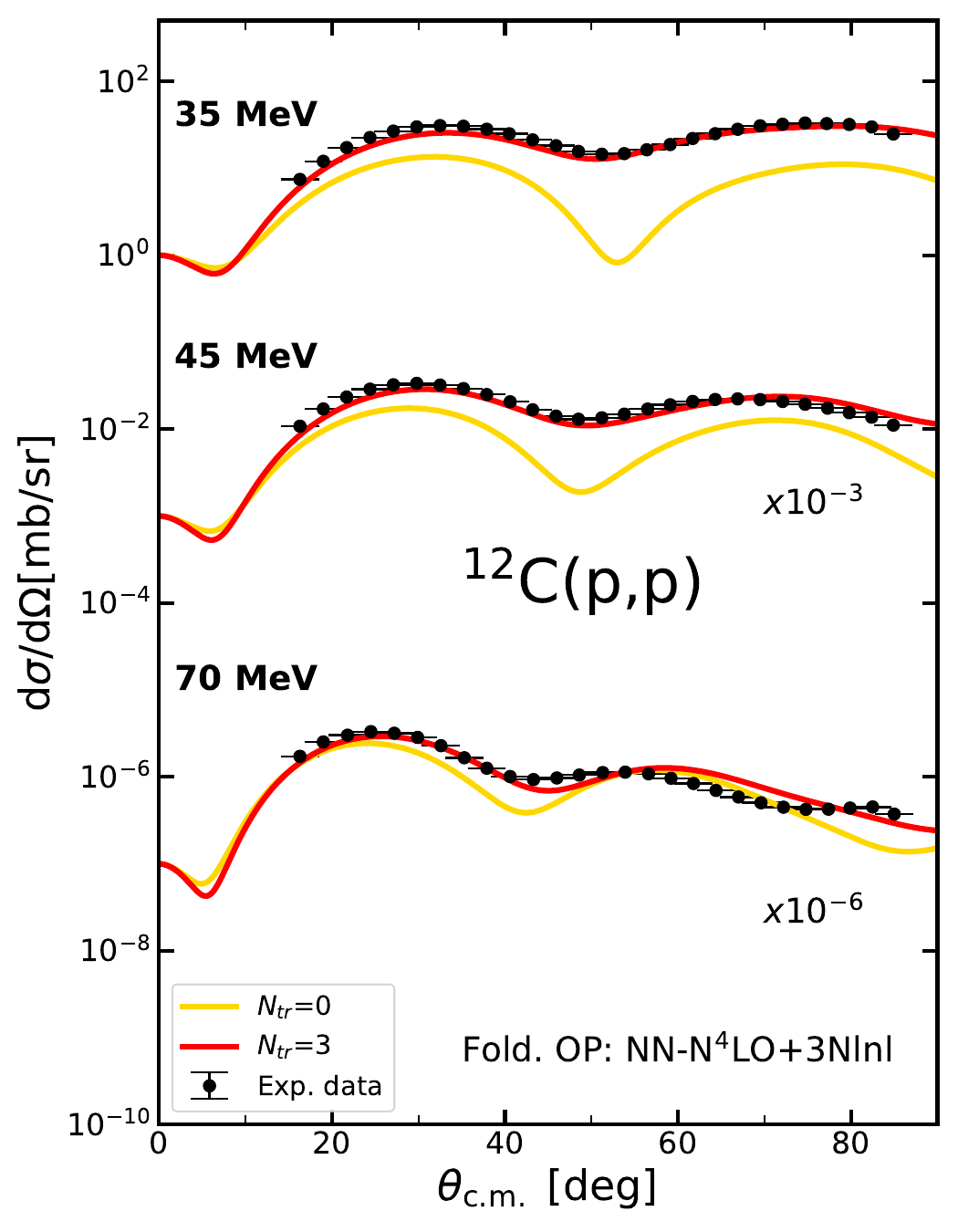}
\caption{\label{figp_12c} 
Differential cross sections as functions of the center-of-mass scattering angle for the reaction $^{12}$C($p,p$) at 35, 45 MeV (scaled by a factor $10^{-3}$) and 70 MeV (scaled by a factor $10^{-6}$). The folded optical potential is calculated with the NN-N$^4$LO+3Nlnl chiral interaction. The results are shown only for the orders $N_{{\rm tr}}=0$ (yellow lines) and $N_{{\rm tr}}=3$3 (red lines).
Experimental data (with error bars) are taken from Ref.~\cite{PhysRevC.33.40,IEIRI1987253}. }
\end{figure}

\begin{figure}
\centering
\includegraphics[scale=0.65, angle=0.0]{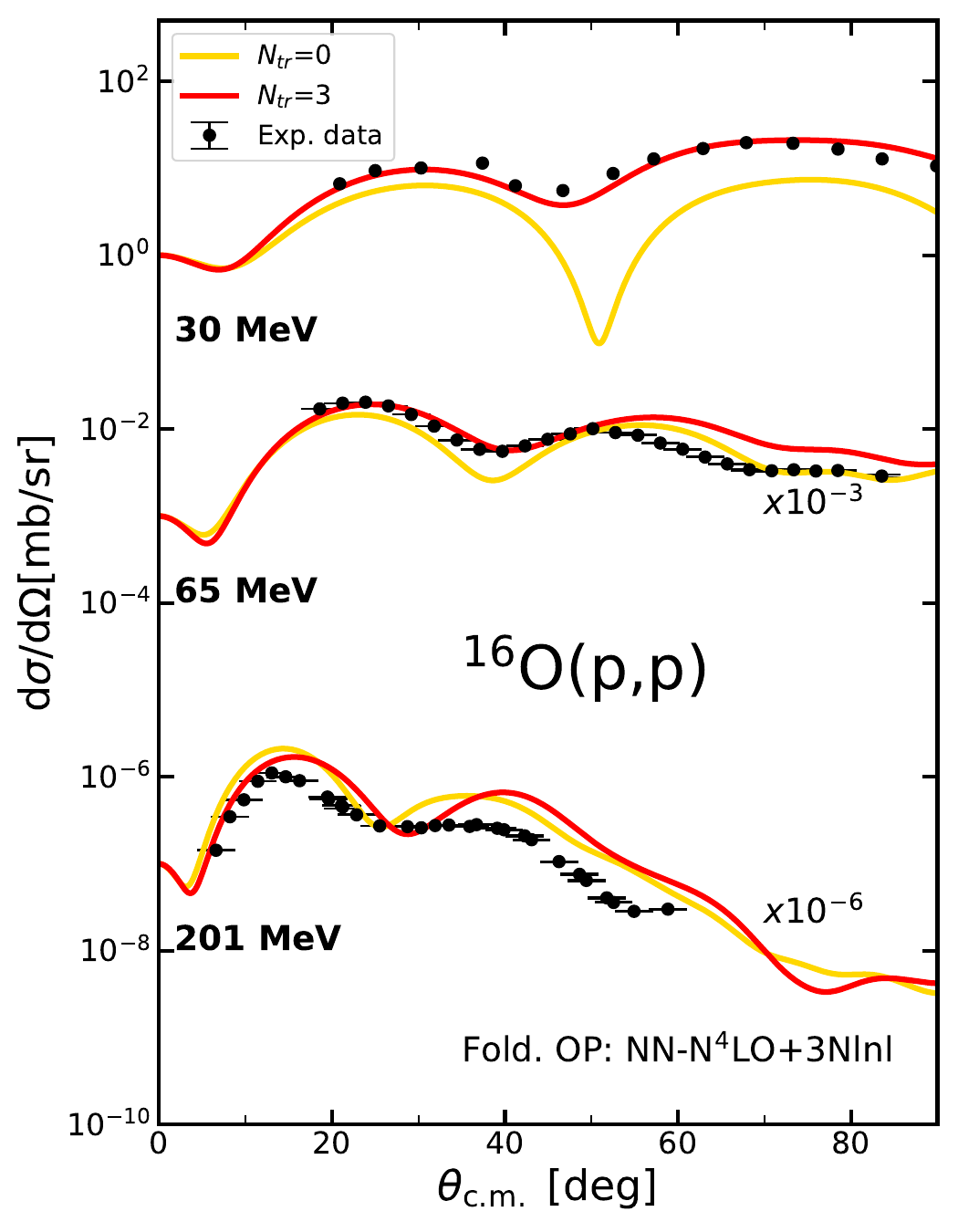}
\caption{\label{figp_16o} 
Differential cross sections as functions of the center-of-mass scattering angle for the reaction $^{16}$O($p,p$) at 30, 65 MeV (scaled by a factor $10^{-3}$) and 201 MeV (scaled by a factor $10^{-6}$). The folded optical potential is calculated with the NN-N$^4$LO+3Nlnl chiral interaction. The results are shown only for the orders $N_{{\rm tr}}=0$ (yellow lines) and $N_{{\rm tr}}=3$3 (red lines).
Experimental data (with error bars where available) are taken from Ref.~\cite{BUNKER1971378, KARBAN1969548}. }
\end{figure}

Conceptually, similar results are obtained for elastic proton scattering.
Differential cross sections for elastic proton scattering on $^{12}$C and $^{16}$O are shown in Figs.~\ref{figp_12c} and \ref{figp_16o}, respectively. In each figure the theoretical results truncated at $N_{{\rm tr}}=0$ and $N_{{\rm tr}}=3$ are displayed at three increasing incident proton energies and compared with experimental data. 
The comparison highlights the overall agreement between our results and data in both magnitude and angular dependence, as well as the important contribution of the proposed phenomenological correction to this agreement.

We observe two distinct behaviours. On the one hand, our phenomenological correction allows us to significantly improve the description of the experimental differential cross sections at low energies, thus allowing us to go beyond the single-scattering approximation. On the other hand, the contribution of the proposed corrections progressively decreases as energy increases. This is consistent with the basic assumption of the dominance of single-scattering processes. Corrections are larger and crucial to describe data for projectile energies close to 30 MeV, and less significant for energies close to and higher than 70 MeV. 

\begin{figure}
\centering
\includegraphics[scale=0.65, angle=0.0]{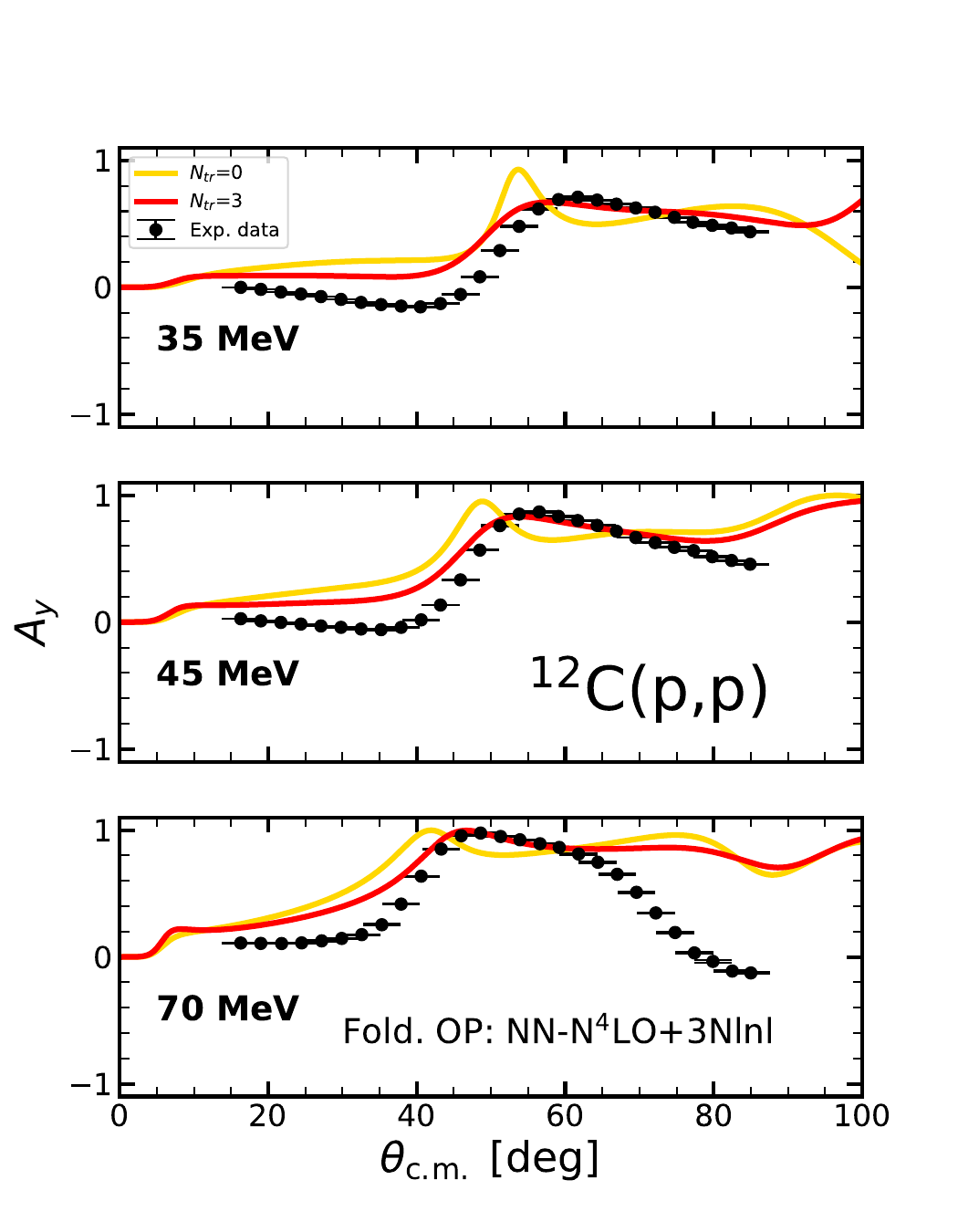}
\caption{\label{figay_12c} Analyzing power as functions of the center-of-mass scattering angle for the reaction $^{12}$C($p,p$) at 30, 45 MeV and 75 MeV. The folded optical potential is calculated with the NN-N$^4$LO+3Nlnl chiral interaction. The results are shown only for the orders $N_{{\rm tr}}=0$ (yellow lines) and $N_{{\rm tr}}=3$3 (red lines). Experimental data are taken from Ref.~\cite{IEIRI1987253}. }
\end{figure}

The behaviour of the truncation scheme for spin observables is illustrated in Fig.~\ref{figay_12c}, where the analyzing power $A_y$ for the elastic reaction $^{12}$C($p,p$) is shown at projectile energies of 35, 45, and 70 MeV. The comparison between the leading-order $(N_{\rm tr}=0$) and higher-order  ($N_{\rm tr}=3$) calculations highlights the impact of multiple scattering corrections on polarization observables.

As observed in the figure, the inclusion of higher-order terms leads to a slight modification of the angular dependence of $A_y$, in particular, improving the overall trend and the position of extrema. The higher-order calculation generally provides a smoother and more physically consistent behaviour compared to the leading-order result, which tends to overestimate the amplitude of oscillations and exhibits less realistic angular structures. At the same time, a quantitative reproduction of the experimental data is not achieved across the full angular range. This should not be interpreted as a limitation of the present truncation scheme, but rather as a reflection of the intrinsic sensitivity of polarization observables to ingredients that go beyond the present framework, such as detailed spin-dependent components of the interaction, higher-order correlations, and possible deficiencies in the underlying nuclear structure input. In particular, analyzing powers are well known to probe subtle interference effects and are therefore significantly more demanding than differential cross sections.

In this context, the role of the present approach is not to provide a fully quantitative description of spin observables, but rather to assess the stability and consistency of the proposed modification of the multiple scattering expansion. From this perspective, the observed convergence between $N_{\rm tr}=0$ and  $N_{\rm tr}=3$, together with the improved qualitative agreement with the data, supports the robustness of the truncation scheme and confirms that higher-order contributions play a non-negligible role also in spin-dependent observables.

A similar analysis can be carried out for the $^{16}$O($p,p$) reaction, as shown in Fig.~\ref{figp_16o} for projectile energies of 30, 65, and 201 MeV. 

\begin{figure}
\centering
\includegraphics[scale=0.65, angle=0.0]{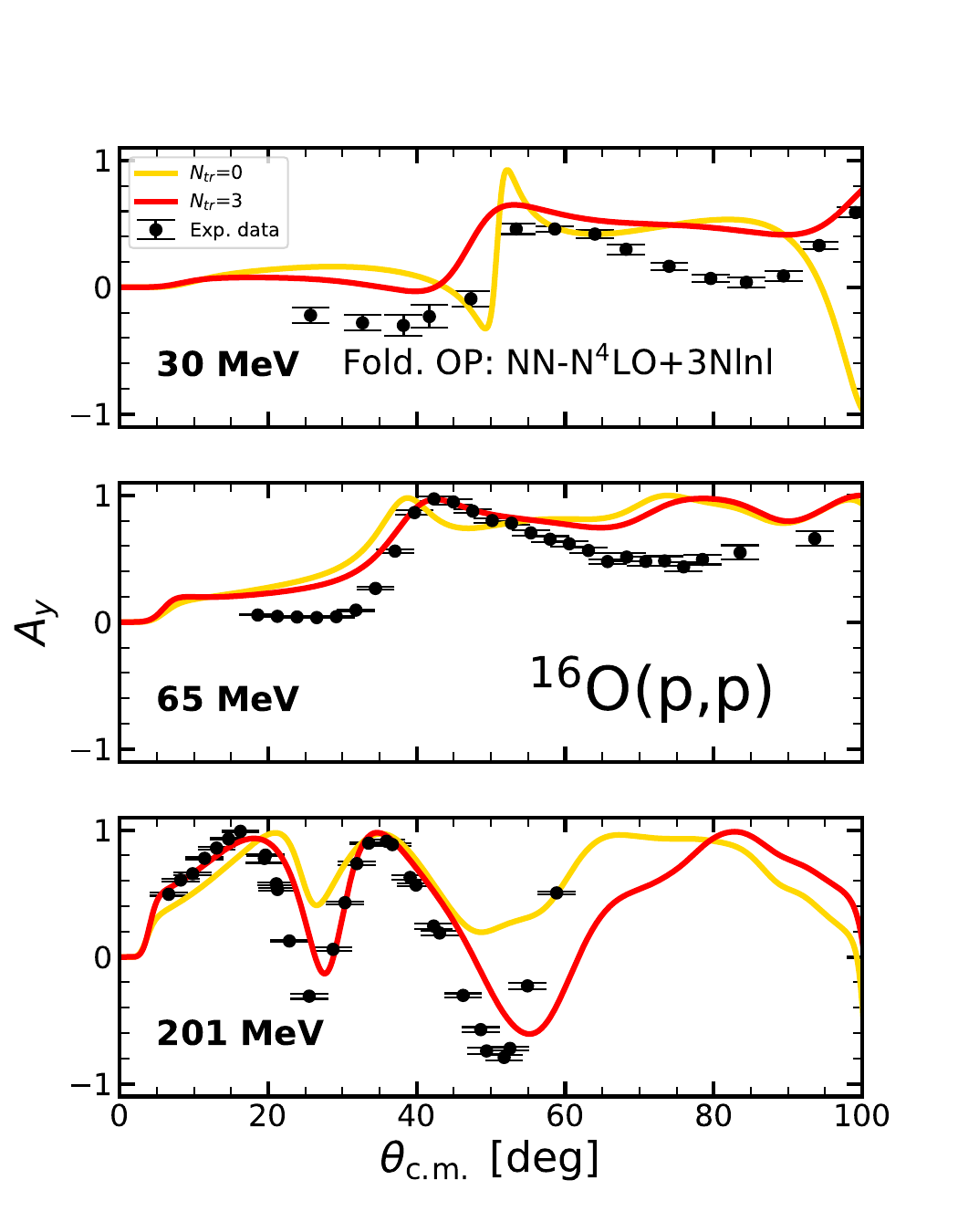}
\caption{\label{figay_16o} Analyzing power as functions of the center-of-mass scattering angle for the reaction $^{16}$O($p,p$) at 30, 65 MeV and 201 MeV. The folded optical potential is calculated with the NN-N$^4$LO+3Nlnl chiral interaction. The results are shown only for the orders $N_{{\rm tr}}=0$ (yellow lines) and $N_{{\rm tr}}=3$3 (red lines). Experimental data are (with error bars where available) taken from Ref.~\cite{PhysRev.167.915,PhysRevC.8.1045}. }
\end{figure}

Also in this case, the comparison between the leading-order ($N_{\rm tr}=0$) and higher-order ($N_{\rm tr}=3$) results demonstrates the systematic impact of multiple scattering corrections on the analyzing power $A_y$.

As visible in the figure, the inclusion of higher-order terms leads to a noticeable improvement in the overall angular behaviour, particularly in the intermediate angular region, where the structure of the observable is more pronounced. The higher-order calculation tends to moderate the oscillatory behaviour of the leading-order result and provides a more stable and physically reasonable description of the observable across different energies.

Nevertheless, significant discrepancies with the experimental data remain, especially at lower energies and in specific angular regions. As in the $^{12}$C case, this should not be regarded as a failure of the model. Instead, it reflects the well-known difficulty of describing polarization observables within microscopic optical potential approaches, particularly when based on interactions and densities not explicitly optimized for spin-dependent observables. The sensitivity of $A_y$ to fine details of the spin–orbit interaction, as well as to medium and correlation effects beyond the present level of approximations, makes a quantitative agreement particularly challenging.

It is also worth noting that the discrepancies tend to decrease with increasing projectile energy, where the underlying assumptions of the reaction framework are better justified. At 201 MeV the higher-order calculation captures several qualitative features of the data, although differences in the amplitude and phase of the oscillations persist.

Overall, these results confirm that the present truncation scheme provides a consistent and controlled framework for incorporating higher-order effects, even in the case of demanding spin observables. The remaining discrepancies should be interpreted as an indication of missing physical ingredients beyond the current level of description rather than as a limitation of the expansion itself.

\section{Conclusions}

In this work we have introduced a phenomenological correction scheme designed to supplement microscopic optical potentials with missing medium and higher-order scattering contributions. By incorporating an additional energy-dependent term, our approach captures effects associated with dispersive couplings, multi-step scattering, and dynamical correlations that are not fully represented in conventional microscopic formulations. The corrected potentials provide a significantly improved description of elastic scattering observables, reducing long-standing discrepancies between theory and experiment while retaining a clear microscopic foundation.

The results presented here underscore the promise of combining microscopic theory with judicious phenomenological input. This hybrid strategy preserves the predictive character of microscopic models, particularly their connection to underlying nucleon–nucleon forces and nuclear structure, while enhancing flexibility and accuracy in practical applications. The encouraging agreement with experimental data across different target nuclei suggests that the correction scheme offers a robust and transferable framework, potentially applicable well beyond the test cases considered.

Looking forward, the methodology can be extended to reactions involving exotic nuclei, where reliable optical potentials are essential yet purely phenomenological models often fail due to lack of experimental guidance. Moreover, the correction scheme offers a natural bridge toward dispersive optical models and other unified descriptions of nuclear reactions and structure. With further refinement and systematic benchmarking, the proposed framework may provide an essential tool for next-generation nuclear reaction studies, with impact ranging from fundamental tests of nuclear interactions to applications in astrophysics and nuclear technology.

\section{Funding}

This work used the DiRAC Data Intensive service (DIaL3) at the University of Leicester, managed by the University of Leicester Research Computing Service on
behalf of the STFC DiRAC HPC Facility (www.dirac.ac.uk). The DiRAC service at Leicester was funded by BEIS, UKRI and STFC capital funding and STFC operations
grants. DiRAC is part of the UKRI Digital Research Infrastructure. This work used the DiRAC Complexity system, operated by the University of Leicester IT Services,
which forms part of the STFC DiRAC HPC Facility (www.dirac.ac.uk ). This equipment is funded by BIS National E-Infrastructure capital grant ST/K000373/1 and STFC
DiRAC Operations grant ST/K0003259/1. DiRAC is part of the National e-Infrastructure.
This work was supported in part by the Deutsche Forschungsgemeinschaft (DFG) through the Cluster of Excellence "Precision Physics, Fundamental Interactions,
and Structure of Matter'" (Project ID 390831469).
P. N. acknowledges support from the NSERC Grant No. SAPIN-2022-00019. TRIUMF receives federal funding via a contribution agreement with the National
Research Council of Canada. Computing support also came from an INCITE Award on the Summit and Frontier supercomputers of the Oak Ridge Leadership Computing
Facility (OLCF) at ORNL and from the Digital Research Alliance of Canada.


\providecommand{\newblock}{}

\end{document}